%Paper: hep-ph/9404210
%From: FARAGGI@sns.ias.edu
%Date: 04 Apr 1994 09:39:11 -0400 (EDT)

%
% Printing instructions:
%       This paper needs the macro packages phyzzx.tex and tables.tex
%       The tables should be stripped off and printed separately.
%
%
\input phyzzx
\tolerance=1000
\sequentialequations
\def\rl{\rightline}

\def\t1{{\tilde 1}}

\def\AEF{A.E.Faraggi}
\def\DVN{D.V.Nanopoulos}

\def\NPB#1#2#3{Nucl.Phys.B {\bf#1} (19#2) #3}
\def\PLB#1#2#3{Phys.Lett.B {\bf#1} (19#2) #3}

\def\PRL#1#2#3{Phys.Rev.Lett. {\bf#1} (19#2) #3}

\REF\ALTARELLI{G.Altarelli, talk at this meeting.}
\REF\UP{D.Gross and P.Mende, \PLB{197}{87}{129};
        D.Amati, M.Ciafaloni and G.Veneziano, \PLB{216}{89}{41}.}
\REF\FFF{I.Antoniadis, C.Bachas, and C.Kounnas, \NPB{289}{87}{87};
H.Kawai, D.C.Lewellen, and S.H.-H.Tye, \NPB{288}{87}{1}.}
\REF\HETE{D.J.Gross, J.A.Harvey, J.A.Martinec and R.Rohm,
                        \PRL{54}{85}{502}; \NPB{256}{86}{253}.}
\REF\KLN{S.Kalara, J.Lopez and D.V.Nanopoulos, \NPB{353}{91}{650}.}
\REF\REVAMP{I.Antoniadis, J.Ellis, J.Hagelin, and \DVN, \PLB{231}{89}{65}.}
\REF\FM{\AEF, \NPB{407}{93}{57}, hep-ph/9210256; IASSNS--HEP--93/77, Phys.
               Lett. B, in press, hep-ph/9311312.}
\REF\DHVW{L.Dixon, J.A.Harvey, C.Vafa and E.Witten,
\NPB{274}{86}{285}.}
\REF\FNY{\AEF, D.V.Nanopoulos and K.Yuan, \NPB{335}{90}{347}.}
\REF\EU{\AEF,  \PLB{278}{92}{131}.}
\REF\TOP{\AEF, \PLB{274}{92}{47}.}
\REF\SLM{\AEF, \NPB{387}{92}{239}, hep-th/9208024.}
\REF\DSW{M.Dine, N.Seiberg and E.Witten, Nucl.Phys.{\bf B289} (1987) 585.}
\REF\NRT{\AEF, \NPB{403}{93}{101}, hep-th/9208023.}
\REF\CKM{\AEF~ and E.Halyo, \PLB{307}{93}{305}, hep-ph/9301261;
         WIS--93/35/APR--PH, Nucl. Phys. {\bf B}, in press, hep-ph/9306235.}
\REF\GCU{\AEF, \PLB{302}{93}{202}, hep-ph/9301268.}
\REF\SUSYX{\AEF~ and E.Halyo, IASSNS--HEP--94/17, to appear.}

\singlespace
\rl{IASSNS--HEP--94/21}
%\rl{\today}
\rl{March 1994}
\rl{T}
%\normalspace
\baselineskip=18pt
\smallskip
\centerline{\bf{Deriving The Standard Model From Superstring Theory
{\footnote*{\it Invited talk presented at the first international conference
  on phenomenology of unification from present to future, Rome, March 1994.}}
}}
\medskip
\centerline{Alon E. Faraggi
{\footnote*{Work supported by an SSC fellowship.
                 e--mail address: faraggi@sns.ias.edu}}
}
\centerline {School of Natural Science, Institute for Advanced Study}
\centerline {Olden Lane, Princeton, NJ 08540}
\bigskip
\centerline{ABSTRACT}
\baselineskip=18pt
I outline a program to derive the Standard Model directly from
superstring theory. I present a class of three generation superstring
standard--like models in the free fermionic formulation. I discuss
some phenomenological properties of these models. In particular these
models suggest an explanation for the top quark mass hierarchy.
A numerical estimate yielded
$m_t\sim175-180~GeV$. The general texture of fermion mass matrices
was obtained from analysis of nonrenormalizable terms up to order $N=8$.
I argue that the realistic
features of these models are due to the underlying $Z_2\times Z_2$ orbifold,
with standard embedding,
at the free fermionic point in toroidal compactification space.

\medskip
\baselineskip=20pt
%\singlespace
%\vskip 0.5cm
\nopagenumbers
\pageno=0
\endpage
%\normalspace
\pagenumbers

%\centerline{\bf 1. Introduction}

Superstring theory is the only known
framework for the consistent unification of quantum gravity with the
gauge interactions. The idea of gauge unification, and its big
desert scenario, is supported by calculation of $\sin^2\theta_W$,
$m_b/m_\tau$ and the recent observation of only three light left--handed
neutrinos at LEP. Furthermore, LEP precision data, in general, support
the validity of supersymmetric unified theories while they severely
constrain theories that perturb the Standard Model very strongly,
like Technicolor or composite models$^{\ALTARELLI)}$.
Ultimately, the nature of
the electroweak symmetry breaking sector will be decided by future
accelerators.

Despite the success of the Standard Model and the qualitative support for
unified gauge theories, point field theories are in general
incomplete. First, there are still too many parameters and the choice
of unifying group is arbitrary. Second, how does nature choose to have
three generations, and the mass hierarchy among them. In particular
the top--bottom mass hierarchy. Finally, gravity and point
quantum field theories are incompatible. In the context
of unified theories the solution to these problems must come
from a more fundamental Planck scale theory.
Furthermore,
suppose that the Higgs sector of the Standard Model is strongly
interacting or that the quark and leptons are composite.
In this case we would have one point quantum field theory replacing the
current one and we would continue our long march to the Planck scale,
where gravitational interactions cannot be neglected. Eventually we will
be faced with the need to reconcile between gravity and quantum mechanics.
String theory does that in a very mild and natural way. The fundamental
conceptual modification in string theory might be embodied in the modification
of the uncertainty principle$^{\UP)}$.
In string theory the uncertainty principle
receives an additional term that depends on the energy scale over $M_{Pl}$.
Thus, as long as we are confined to low energies
the additional term is small and point quantum field theories are a
good approximation. However, when we get near the Planck the additional
term gets large and point particle approximation breaks down.

Superstring theory provides us with a unique framework to study
how the Planck scale
may determine the parameters of the Standard Model. Keeping in mind
the experimental observations that support the validity of the
Standard Model up to a very large scale,
it makes sense to ask whether it is possible to connect between superstring
theory and the highly successful Standard Model. The aim of the work
that I describe in this talk is to achieve precisely that. Using the
free fermionic model building rules$^{\FFF)}$ we ask whether it is possible
to construct a ``realistic'' superstring standard--like model?
A realistic model must posses the following properties.
First, the gauge group is $SU(3)\times SU(2)\times U(1)^n\times hidden$.
Second, the massless spectrum must have three chiral generations
and a pair of Higgs doublets that can produce a realistic fermion mass
spectrum. Finally, we require $N=1$ space--time supersymmetry. This
last requirement ensures vanishing of the cosmological constant.
However, nonsupersymmetric models, and
nonsupersymmetric models in which some electroweak doublets
transform under a non--Abelian hidden gauge group, can be constructed.
In these models, in general, the cosmological constant does not vanish.

Besides being formulated in four space--time dimensions,
what is special about the free fermionic formulation? The heterotic
string in ten space--time dimensions is more or less unique. However,
when we compactify to four dimensions this uniqueness is lost.
The free fermionic formulation is formulated at a highly symmetric point
in the compactification space. It is an exact Conformal Field Theory
and we can use the CFT calculational tools to calculate Yukawa couplings, etc.
Finally, in free fermionic models
one can naturally obtain three generations with standard $SO(10)$ embedding.

Several properties are required from a heterotic string theory$^{\HETE)}$:
conformal invariance, modular invariance and world--sheet supersymmetry.
In the free fermionic formulation of the heterotic string all the
degrees of freedom that are needed to cancel the conformal anomaly are
represented in terms of free fermions propagating on the string
world--sheet. Under parallel transport around a noncontractible loop
the fermionic states pick up a phase. A model in this construction
is defined by a set of basis vectors of boundary conditions
for all world--sheet fermions. The basis vectors are constrained
by modular invariance and world--sheet supersymmetry and span a finite
additive group $\Xi$. For every vector in the additive group
correspond a sector in the string Hilbert space. The physical spectrum
is obtained by applying the generalized GSO projections.
The low energy effective
field theory is obtained by S--matrix elements between external states.
The Yukawa couplings and higher order terms in the superpotential
are obtained by calculating correlators between vertex operators$^{\KLN)}$. For
a correlator to be nonvanishing all the symmetries of the model must
be conserved. Thus, the boundary condition vectors completely determine
the phenomenology of the models.

The first five vectors in the basis that generate the standard--like
models consist of the NAHE{\footnote*{This set was first
constructed$^{\REVAMP)}$
by Nanopoulos, Antoniadis, Hagelin and Ellis  (NAHE)
in the construction
of  the flipped $SU(5)$.  {\it nahe}=pretty, in
Hebrew.}} set, $\{{\bf 1},S,b_1,b_2,b_3\}$,
and are common to all the realistic
free fermionic models. The gauge group after the NAHE set is
$SO(10)\times SO(6)^3\times E_8$ with $N=1$ space--time supersymmetry,
and 48 spinorial $16$ of $SO(10)$, sixteen from each sector
$b_1$, $b_2$ and $b_3$. The NAHE set divides the internal world--sheet
fermions in the following way: ${\bar\phi}^{1,\cdots,8}$ generate the
hidden $E_8$ gauge group, ${\bar\psi}^{1,\cdots,5}$ generate the $SO(10)$
gauge group, and $\{{\bar y}^{3,\cdots,6},{\bar\eta}^1\}$,
$\{{\bar y}^1,{\bar y}^2,{\bar\omega}^5,{\bar\omega}^6,{\bar\eta}^2\}$,
$\{{\bar\omega}^{1,\cdots,4},{\bar\eta}^3\}$ generate the three horizontal
$SO(6)^3$ symmetries. The left--moving $\{y,\omega\}$ states are divided
to $\{{y}^{3,\cdots,6}\}$,
$\{{y}^1,{y}^2,{\omega}^5,{\omega}^6\}$,
$\{{\omega}^{1,\cdots,4}\}$ and $\chi^{12}$, $\chi^{34}$, $\chi^{56}$
generate the left--moving $N=2$ world--sheet supersymmetry.

The internal fermionic states $\{y,\omega\vert{\bar y},{\bar\omega}\}$
correspond to the six left--moving and six right--moving compactified
dimensions in a geometric formulation. This correspondence is illustrated
by adding the vector with periodic boundary conditions for the set
$\{{{\bar\psi}^{1,\cdots,5}},
{{\bar\eta}^{1,2,3}}\}$
to the NAHE set$^{\FM)}$. This extends the gauge group to
$E_6\times U(1)^2\times E_8\times SO(4)^3$ with $N=1$ space--time
supersymmetry
and twenty four chiral $27$ of $E_6$. The same model is generated in the
orbifold language$^{\DHVW)}$
by moding out an $SO(12)$ lattice by a $Z_2\times{Z_2}$
discrete symmetry with standard embedding. The $SO(12)$ lattice
is obtained for special values of the metric and antisymmetric tensor
and at the self dual point in compactification space. The metric is the Cartan
matrix of $SO(12)$ and the antisymmetric tensor is given by $b_{ij}=g_{ij}$
for $i>j$. The sectors $b_1$, $b_2$ and $b_3$ correspond to
the three twisted sectors in the orbifold models and the Neveu--Schwarz sector
corresponds to the untwisted sector.
In the construction of
the standard--like models beyond the NAHE set, the assignment
of boundary conditions to the set of internal fermions
$\{y,\omega\vert{\bar y},{\bar\omega}\}$ determines many of the
properties of the low energy spectrum, such as the number of
generations, the presence of Higgs doublets, Yukawa couplings, etc.

The standard--like models are constructed by adding three additional
vectors to the NAHE set$^{\FNY,\EU,\TOP,\SLM)}$.
One example is presented in the table, where only the boundary conditions
of the ``compactified space" are shown. In the gauge sector
$\alpha,\beta\{{{\bar\psi}^{1,\cdots,5}},
{{\bar\eta}^{1,2,3}},{\bar\phi}^{1,\cdots,8}\}=\{1^3,0^5,1^4,0^4\}$
and $\gamma\{{{\bar\psi}^{1,\cdots,5}},
{{\bar\eta}^{1,2,3}},{\bar\phi}^{1,\cdots,8}\}=\{{1\over2}^9,0,1^2,
{1\over2}^3,0\}$ break the symmetry to $SU(3)\times SU(2)\times
U(1)_{B-L}\times U(1)_{T_{3_R}}\times SU(5)_h\times SU(3)_h\times U(1)^2$.
Three additional vectors are needed to reduce the
number of generations to one generation from each sector $b_1$, $b_2$
and $b_3$. Each generation has horizontal symmetries that constrain
the allowed interactions. Each generation has two gauged $U(1)$
symmetries, $U(1)_{R_j}$ and $U(1)_{R_{j+3}}$. For every right--moving
$U(1)$ symmetry there is a corresponding left--moving global $U(1)$ symmetry,
$U(1)_{L_j}$ and $U(1)_{L_{j+3}}$. Finally, each generation has two Ising
model operators that are obtained by pairing a left--moving real fermion
with a right--moving real fermion.
\input tables.tex
\smallskip
{{\it Table 1.} A three generations ${SU(3)\times SU(2)\times U(1)^2}$
                model$^{\TOP)}$.
\vskip .75mm
{\hfill
{\begintable
\  \ \|\
${y^3y^6}$,  ${y^4{\bar y}^4}$, ${y^5{\bar y}^5}$,
${{\bar y}^3{\bar y}^6}$
\ \|\ ${y^1\omega^6}$,  ${y^2{\bar y}^2}$,
${\omega^5{\bar\omega}^5}$,
${{\bar y}^1{\bar\omega}^6}$
\ \|\ ${\omega^1{\omega}^3}$,  ${\omega^2{\bar\omega}^2}$,
${\omega^4{\bar\omega}^4}$,  ${{\bar\omega}^1{\bar\omega}^3}$  \crthick
${\alpha}$ \|
1, ~~~1, ~~~~1, ~~~~0 \|
1, ~~~1, ~~~~1, ~~~~0 \|
1, ~~~1, ~~~~1, ~~~~0 \nr
${\beta}$ \|
0, ~~~1, ~~~~0, ~~~~1 \|
0, ~~~1, ~~~~0, ~~~~1 \|
1, ~~~0, ~~~~0, ~~~~0 \nr
${\gamma}$ \|
0, ~~~0, ~~~~1, ~~~~1 \|\
1, ~~~0, ~~~~0, ~~~~0 \|
0, ~~~1, ~~~~0, ~~~~1
\endtable}}
Higgs doublets in the standard--like models are obtained from two
distinct sectors. The first type are obtained from the Neveu--Schwarz sector,
which produces three pairs of electroweak doublets. Each pair can couple
at tree level only to the states from the sector $b_j$. There is a
stringy doublet--triplet splitting mechanism that projects out the
color triplets and leaves the electroweak doublets in the spectrum.
Thus, the superstring standard--like models resolve the GUT hierarchy
problem. The second type of Higgs doublets are obtained from
the vector combination $b_1+b_2+\alpha+\beta$. The states in this
sector are obtained by acting on the vacuum with a single fermionic
oscillator and transform only under the observable sector.

The cubic level Yukawa couplings for the quarks and leptons are
determined by the boundary conditions in the vector $\gamma$
according to the following rule$^{\SLM)}$
$$\eqalignno{\Delta_j&=
\vert\gamma(U(1)_{\ell_{j+3}})-\gamma(U(1)_{r_{j+3}})\vert=0,1
{\hskip 1cm}(j=1,2,3)&(3a)\cr
\Delta_j&=0\rightarrow d_jQ_jh_j+e_jL_jh_j;{\hskip .2cm}
\Delta_j=1\rightarrow u_jQ_j{\bar h}_j+N_jL_j{\bar h}_j,&(3b,c)\cr}$$ where
$\gamma(U(1)_{R_{j+3}})$, $\gamma(U(1)_{\ell_{j+3}})$ are the boundary
conditions of the world--sheet fermionic currents that generate the
$U(1)_{R_{j+3}}$, $U(1)_{\ell_{j+3}}$ symmetries.

The superstring standard--like models contain an anomalous $U(1)$
gauge symmetry. The anomalous $U(1)$ generates a Fayet--Iliopoulos term
by the VEV of the dilaton field that breaks supersymmetry and destabilizes
the vacuum$^{\DSW)}$.
Supersymmetry is restored by giving VEVs to standard model
singlets in the massless spectrum of the superstring models. However,
as the charge of these singlets must have $Q_A<0$ to cancel the anomalous
$U(1)$ D--term equation, in  many models a phenomenologically realistic
solution does not exist. In fact a very restricted class of standard--like
models with $\Delta_j=1$ for $j=1,2,3$, were found to admit a solution
to the F and D flatness constraints. Consequently, the only models that
were found to admit a solution are models which have tree level
Yukawa couplings only for $+{2\over3}$ charged quarks.

This result suggests an explanation for the top quark mass hierarchy
relative to the lighter quarks and leptons. At the cubic level only the
top quark gets a mass term and the mass terms for the lighter
quarks and leptons are obtained from nonrenormalizable terms.
To study this scenario we have to examine the nonrenormalizable
contributions to the doublet Higgs mass matrix and to the fermion mass
matrices$^{\NRT,\CKM)}$.

At the cubic level there are two pairs of electroweak doublets.
At the nonrenormalizable level one additional pair receives a superheavy
mass and one pair remains light to give masses to the fermions at
the electroweak scale. Requiring F--flatness imposes that the light
Higgs representations are ${\bar h}_1$ or ${\bar h}_2$ and $h_{45}$.

The nonrenormalizable fermion mass terms of order $N$ are of the form
$cgf_if_jh\phi^{^{N-3}}$ or
$cgf_if_j{\bar h}\phi^{^{N-3}}$, where $c$ is a
calculable coefficient, $g$ is the gauge coupling at the unification
scale,  $f_i$, $f_j$ are the fermions from
the sectors $b_1$, $b_2$ and $b_3$, $h$ and ${\bar h}$ are the light
Higgs doublets, and $\phi^{N-3}$ is a string of standard model singlets
that get a VEV and produce a suppression factor
${({{\langle\phi\rangle}/{M}})^{^{N-3}}}$ relative to the cubic
level terms. Several scales contribute to the generalized VEVs. The
leading one is the scale of VEVs that are used to cancel the anomalous
D--term equation. The next scale is generated by Hidden sector
condensates. Finally, there is a scale which is related to the breaking
of $U(1)_{Z^\prime}$, $\Lambda_{Z^\prime}$. Examination of the higher
order nonrenormalizable terms reveals that $\Lambda_{Z^\prime}$ has
to be suppressed relative to the other two scales.

At the cubic level only the top quark gets a nonvanishing mass term.
Therefore only the top quark mass is characterized by the electroweak
scale. The remaining quarks and leptons obtain their mass terms from
nonrenormalizable terms. The cubic and nonrenormalizable terms in the
superpotential are obtained by calculating correlators between the vertex
operators. The top quark Yukawa coupling is generically given by
$$g\sqrt2\eqno(1)$$
where $g$ is the gauge coupling at the unification scale. In
the model of Ref. [\TOP], bottom quark and tau lepton mass terms
are obtained at the quartic order,
$$W_4=\{{d_{L_1}^c}Q_1h_{45}^\prime\Phi_1+{e_{L_1}^c}L_1h_{45}^\prime\Phi_1+
{d_{L_2}^c}Q_2h_{45}^\prime{\bar\Phi}_2
+{e_{L_2}^c}L_2h_{45}^\prime{\bar\Phi}_2\}.\eqno(2)$$
The VEVs of $\Phi$ are obtained from the cancelation of the anomalous
D--term equation. The coefficient of the quartic order mass terms
were calculated by calculating the quartic order correlators and the
one dimensional integral was evaluated numerically. Thus after
inserting the VEV of ${\bar\Phi}_2$ the effective bottom quark and tau lepton
Yukawa couplings are given by$^{\TOP)}$,
$$\lambda_b=\lambda_\tau=0.35g^3.\eqno(3)$$
They are suppressed relative to the top Yukawa by
$${{\lambda_b}\over{\lambda_t}}=
{{0.35g^3}\over{g\sqrt2}}\sim{1\over8}.\eqno(4)$$
To evaluate the top quark mass, the three Yukawa couplings are run
to the low energy scale by using the MSSM RGEs. The bottom mass is
then used to calculate $\tan\beta$ and the top quark mass is
found to be$^{\TOP)}$,
$$m_t\sim175-180GeV.\eqno(5)$$
The fact that the top Yukawa is found near a fixed point suggests that this
is in fact a good prediction of the superstring standard--like models.
By varying $\lambda_t\sim0.5-1.5$ at the unification
scale, it is found that $\lambda_t$ is always $O(1)$ at the
electroweak scale.

An analysis of fermion mass terms up to order $N=8$ revealed the general
texture of fermion mass matrices in these models. The sectors $b_1$
and $b_2$ produce the two heavy generations and the sector $b_3$
produces the lightest generation. This is due to the horizontal
$U(1)$ charges and because the Higgs pair $h_3$ and ${\bar h}_3$
necessarily get a Planck scale mass$^{\NRT)}$.
The mixing between the generations
is obtained from exchange of states from the sectors $b_j+2\gamma$. The
general texture of the fermion mass matrices in the superstring
standard--like models is of the following form,
$${M_U\sim\left(\matrix{\epsilon,a,b\cr
                    {\tilde a},A,c \cr
                    {\tilde b},{\tilde c},\lambda_t\cr}\right);{\hskip .2cm}
M_D\sim\left(\matrix{\epsilon,d,e\cr
                    {\tilde d},B,f \cr
                    {\tilde e},{\tilde f},C\cr}\right);{\hskip .2cm}
M_E\sim\left(\matrix{\epsilon,g,h\cr
                    {\tilde g},D,i \cr
                    {\tilde h},{\tilde i},E\cr}\right)},$$
where $\epsilon\sim({{\Lambda_{Z^\prime}}/{M}})^2$.
The diagonal terms in capital letters represent leading
terms that are suppressed by singlet VEVs, and
$\lambda_t=O(1)$. The mixing terms are generated by hidden sector states
from the sectors $b_j+2\gamma$ and are represented by small letters. They
are proportional to $({{\langle{TT}\rangle}/{M}^2})$. In Ref.
[\CKM] it was shown that if the states from the sectors $b_j+2\gamma$ obtain
VEVs in the application of the DSW mechanism, then a Cabibbo angle of the
correct order of magnitude can be obtained in the superstring standard--like
models. In later work the analysis was extended to show that reasonable
values for the entire CKM matrix parameters can be obtained for appropriate
flat F and D solutions. Texture zeroes in the fermion mass matrices are
obtained if the VEVs of some states from the sectors $b_j+2\gamma$ vanish.
These texture zeroes are protected by the symmetries of the string models
to all order of nonrenormalizable terms$^{\CKM)}$.

Next, I turn to the problem of gauge coupling
unification in the superstring standard--like models$^{\GCU)}$.
While LEP results indicate that the gauge coupling in the minimal
supersymmetric standard model unify at $10^{16}GeV$, superstring
theory predicts that the unification scale is at $10^{18}GeV$.
The superstring standard--like models may resolve this
problem due to the existence of color triplets and electroweak doublets
from exotic sectors that arise from the additional vectors $\alpha$,
$\beta$ and $\gamma$. These exotic states carry fractional charges and
do not fit into standard $SO(10)$ representations. Therefore, they
contribute less to the evolution of the $U(1)_Y$ beta function than
standard $SO(10)$ multiplets.
The standard--like models predict
$\sin^2\theta_W={3/8}$ at the unification scale due to the
embedding of the weak hypercharge in $SO(10)$. In Ref. [\GCU],
I showed that provided that the additional exotic color triplets and
electroweak doublets exist at the appropriate scales, the
scale of gauge coupling unification is pushed to $10^{18}GeV$, with the
correct value of $\sin^2\theta_W$ at low energies.

Next, I comment on the problem of supersymmetry breaking. In Ref. [\SUSYX]
we address the following question: Given a supersymmetric string
vacuum at the Planck scale, is it possible to obtain hierarchical
supersymmetry breaking in the observable sector? A supersymmetric
string vacuum is obtained by finding solutions to the cubic level
F and D constraints. We take a gauge coupling in agreement with
gauge coupling unification, thus taking a fixed value for the dilaton VEV.
We then investigate the role of nonrenormalizable terms and strong
hidden sector dynamics. The hidden sector contains two non--Abelian
hidden gauge groups, $SU(5)\times SU(3)$, with matter in vector--like
representations. The hidden $SU(3)$ group is broken near the Planck scale.
We analyze the dynamics of the hidden $SU(5)$ group.
The hidden $SU(5)$ matter mass matrix is nonsingular for specific F and D
flat solutions. We find that, in some scenarios, the matter and gaugino
condensates can brake supersymmetry in a hierarchically small scale.

To conclude, the superstring standard--like models contain in their
massless spectrum all the necessary states to obtain realistic
phenomenology. They resolve the problems of proton decay through
dimension four and five operators that are endemic to other
superstring and GUT models. The existence of only three generations with
standard $SO(10)$ embedding is understood to arise naturally from
$Z_2\times Z_2$ twisting at the free fermionic point in compactification
space. Better understanding of the correspondence with other superstring
formulations will provide further insight into the realistic properties
of these models. Finally,
the free fermionic standard-like models
provide a highly constrained and phenomenologically realistic laboratory
to study how the Planck scale may determine the parameters of the
Standard Model.

\baselineskip=12pt
\refout
\end